 \documentclass[prl,aps,twocolumn,showpacs]{revtex4}
\usepackage[dvips]{graphicx}
\usepackage{dcolumn}%
\usepackage{amsmath}%
\usepackage{amsfonts}%
\usepackage{amssymb}
 \usepackage[usenames]{color}
\providecommand{\U}[1]{\protect\rule{.1in}{.1in}}
\newcommand{\be}{\begin{equation}}
\newcommand{\ee}{\end{equation}}
\newcommand{\bea}{\begin{eqnarray}}
\newcommand{\eea}{\end{eqnarray}}
\newcommand{\bt} {\begin{tabular}}
\newcommand{\et} {\end{tabular}}

\newcommand{\ds}{\displaystyle}
\newcommand{\ba} {\begin{array}}
\newcommand{\ea} {\end{array}}
\begin{document}

\title{Temperature dependent charge transport mechanisms in carbon sphere/polymer composites}

\author{Cesar A. Nieves$^1, $ Luis M. Martínez$^1,$ Anamaris Melendez$^1,$ Margarita Ortiz$^2,$ Idalia Ramos$^1,$ Nicholas J. Pinto$^1,$ and Natalya A. Zimbovskaya$^1.$ }
\affiliation
{$^1$Department of Physics and Electronics, University of Puerto 
Rico,  Humacao, Puerto Rico 00791, USA}
\affiliation
{$^2$Department of Chemistry, University of Puerto 
Rico,  Humacao, Puerto Rico 00791, USA}

\begin{abstract}
Carbon spheres (CS) with diameters in the range $2 - 10 \mu m$ were prepared via hydrolysis of a sucrose solution at $200^o C,$ and later annealed in $N_2$ at $800^o C.$ The spheres were highly conducting but difficult to process into thin films or pressed pellets. In our previous work, composite samples of CS and the insulating polymer polyethylene oxide (PEO) were prepared and their charge transport was analyzed in the temperature range $ 80 K < T < 300 K. $ Here, we analyze charge transport in CS coated with a thin polyaniline (PANi) film doped with hydrochloric acid (HCl), in the same temperature range. The goal is to study charge transport in the CS using a conducting polymer (PANi) as a binder and compare with that occurring at CS/PEO. A conductivity maxima was observed in the CS/PEO composite but was absent in CS/PANi. Our data analysis shows that variable range hopping of electrons between polymeric chains in PANi-filled gaps between CS takes on a predominant part in transport through CS/PANi composites, whereas in CS/PEO composites, electrons travel through gaps between CS solely by means of direct tunneling. This difference in transport mechanisms results in different temperature dependences of the conductivity.
 	\end{abstract}

\date{\today}
\maketitle

\section{1. introduction}

Charge transport mechanisms in  condensed matter  
 composites vary depending on the atomic configuration, molecular ordering, purity, crystallinity and temperature of the material, among other factors. Typical mechanisms include ballistic transport, 
 tunneling, activated transport or variable range hopping transport in disordered materials like conducting polymers.  Since composite materials are made from several components whose properties may significantly differ, charge transport in these materials may be simultaneously governed by various transport mechanisms. A better understanding of charge transport at the molecular level is important to tailor material properties for practical applications \cite{1,2,3,4,5,6,7,8,9}. The discovery of conducting polymers in the late 1970’s  \cite{10} has opened new avenues of research in the field of plastic electronics where organic conductors are used as substitutes for traditional metals or semiconductors. Over the past two decades, the field of nanoscience and nanotechnology has focused on carbon based materials like fullerenes \cite{11,12}, carbon nanotubes \cite{13} and graphene \cite{14}. 
  Renewed interest in carbon microspheres for use in high strength carbon based materials and passive electronic components is now making a comeback \cite{15, 16}. Some techniques to make these spheres include chemical vapor deposition, templates, pyrolysis of carbon rich sources, and hydrothermal carbonization of large sugar molecules \cite{17,18,19,20,21,22,23}. An important feature of these spheres is their high conductivity after annealing. They are also chemically very stable under standard atmospheric conditions, and their fabrication is environmentally friendly \cite{19}. 

The carbon spheres are hard and difficult to process into stable films or pellets for practical applications. By forming composites with polyethylene oxide (PEO) we could use the spheres (CS/PEO) in the fabrication of diodes and subsequent half wave rectifier for low frequency signals \cite{24}. In addition, we also analyzed the charge transport in these composites in the temperature range $ 300 K >  T > 80 K $ \cite{9}. Since PEO is an electrically insulating polymer, charge transport in the bulk CS/PEO sample was limited by insulating barriers between adjacent CS. We observed a conductivity $(\sigma) $ maximum around $ 258 K $ and our data analysis showed that tunneling between the CS mostly controls charge transport at high temperatures. At lower temperatures, the contribution from thermally activated electron transport via chains of stuck together carbon spheres becomes comparable with that one originating from tunneling. These two contributions put together are responsible for the conducting maxima. In this paper, we analyze charge transport in carbon spheres coated with a thin conducting polyaniline film doped with hydrochloric acid (HCL) in the same temperature range. The goal was to compare charge transport between the CS using a conducting polymer as the binder with that occurring in CS/PEO. We observe that CS/PANi exhibited higher overall conductivity with a stronger temperature dependence compared to CS/PEO, and is therefore a better composite material to use in  devices and sensors. The conductivity maxima seen in  CS/PEO are absent in CS/PANi.  Charge transport in the latter is via variable range hopping through PANi,  with finite contributions from thermally activated  transport via chains of CS and direct  tunneling between CS. 

\section {2. Experimental}

{\bf 2.a.} {\it Carbon spheres preparation:} Figure 1 shows a schematic diagram of the sphere fabrication process. An aqueous solution of  8.22 g of sucrose dissolved in 30 mL of distilled water was heated at  $ 200^o C$  in an air-tight autoclave for 4 h and then left to cool down to room temperature. This method has been reported in previous work by our group \cite{9, 24}. In summary, the hydrolysis of sucrose produces monosaccharides like glucose, and their subsequent dehydration leads to the formation of polymeric micelles with spherical shapes. Further polymerization leads to spheres with larger diameter \cite{17,18}. The dark brown precipitate in the autoclave was washed with water and ethanol to remove all soluble byproducts from the sucrose decomposition, and dried in an air oven at  $70^o C $ for 12 h.

The as-prepared spheres were then thermally annealed in a tube furnace under dry nitrogen gas flow at  $ 800^o C $ for 1 h, and left to cool down to room temperature. Figure 2(a) shows a scanning electron microscope (SEM) image of these spheres.  The sphere diameters were in the range of  $2 \mu m - 10 \mu m. $ The inset in Figure 2(a) shows the EDX spectrum of the spheres where the reduction of residual O2 leads to a carbon composition of  $99.5 $ at. \% \cite{24}. The spheres were hard, making it difficult to make a pressed pellet or a continuous film for electrical characterization as it would crack even with careful handling. To overcome this, the spheres were mixed with the polymer as explained in sections 2.b and 2.c.

{\bf 2.b.} {\it CS/PEO:} The details of the preparation of the CS/PEO composite were discussed in a previous publication \cite{24}. Briefly, the CS and PEO were mixed in a 1:1 ratio and pressed using a Carver press at 15000 psi. The pressed pellets were 13 mm in diameter and 0.5 mm thick. Figure 2(b) shows a SEM image of the CS/PEO composite together with the corresponding EDX spectrum. The O peak in EDX arises from the presence of PEO and in the composite.

{\bf 2.c.} {\it Polyaniline synthesis and CS coating:} Polyaniline was synthesized via the oxidative polymerization of aniline with ammonium peroxydisulfate (APS) in acidic media at $0^o C $ \cite{25}. In a typical synthesis, CS as prepared above were dispersed in 100 ml of 1M HCl solution in which 20 ml of aniline was dissolved. The mixture was magnetically stirred in a beaker while placed in an ice bath at $0^o C. $ A separate solution was prepared by dissolving 11 g of the oxidant APS in 100 ml of 1M HCl and maintained at $ 0^o C. $ The oxidant was added drop by drop into the aniline/CS solution under constant stirring. The solution began turning deep green following the polymerization of aniline. The solution was left to polymerize for 24 h, then filtered and washed with 1M HCl and dried in air at  $70^o C $ for 48 hours. The formed emeraldine salt of polyaniline (PANi) seems to coat the CS. Figure 2(c) shows a SEM image of CS/PANi composite spheres together with the corresponding EDX spectrum. A closer look at this figure shows a thin PANi polymeric film coating the spheres. The presence of a N peak in the EDX spectrum is the result of PANi coating the spheres. The CS/PANi pressed pellet was prepared using the same procedure described for the CS/PEO in section of 2.b the Experimental Section.

\begin{figure}[t] 
\begin{center}
\includegraphics[width=4.6cm,height=8.8cm,angle=-90]{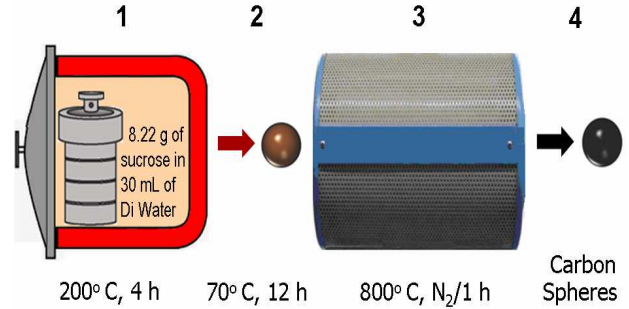} 
\caption{ Schematic of the carbon sphere preparation. 1) Hydrolysis of sucrose and cooling to room temperature in air-tight autoclave. 2) Drying spheres in air oven at $ 70^o$ C  for 12 h. 3) Thermal annealing in a tube furnace at $ 800^o$ C for 1h.  4) The typical yield of carbon spheres after the annealing process was 300 mg.
}
 \label{rateI}
\end{center}\end{figure} 

\begin{figure}[t] 
\begin{center}
 \includegraphics[width=17.5cm,height=7.5cm,angle=-90]{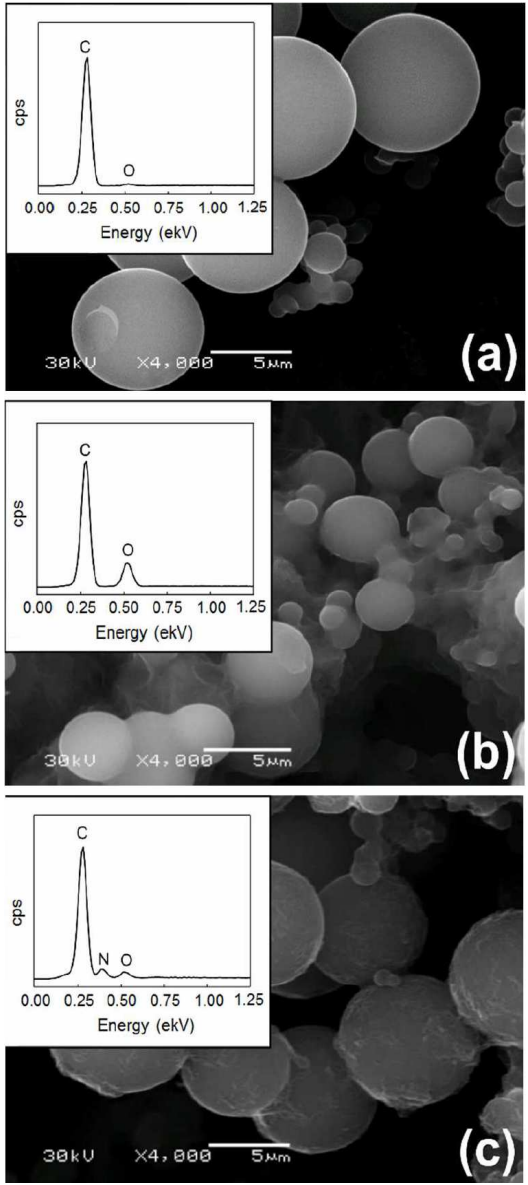}        
\caption{Scanning Electron Microscope images for (a) Annealed CS. (b) CS/PEO composite (c) CS/PANi composite. The insets show the corresponding EDX spectra. The O peak in Figure 2(b) and the N peak in Figure 2(c) arise from the presence of PEO and PANi in the composite.
}
 \label{rateI}
\end{center}\end{figure}

{\bf 2.d.} {\it Electrical characterization:} Rectangular sections of the pressed pellets were cut, and two in- line contacts made with Ag paint and separated by  $ \sim 3 $ mm. The sample was then mounted on the stage of a closed cycle helium refrigerator and the temperature monitored with a Cryo-Con 32B temperature controller. A Keithley electrometer model 6517B was used to supply the voltage and measure the current making sure that the power dissipation was small to avoid Joule heating during the measurement. The measurements were made in a vacuum of 10-2 Torr.  The same method was used for the CS/PANi and CS/PEO composites.

\section {3. Results and Discussion}

{\bf 3a.} {\it Discussion of experimental results:} Figure 3 shows the current-voltage $ (I-V) $ curves of CS/PANi and CS/PEO composites. These results demonstrate that CS/PANi exhibits a higher conductivity than the CS/PEO composite. This is a reasonable result, since PANi is a conducting polymer that bridges the conducting CS lowering the barriers to charge transport between the spheres.

\begin{figure}[t] 
\begin{center}
\includegraphics[width=7cm,height=7cm,angle=-90]{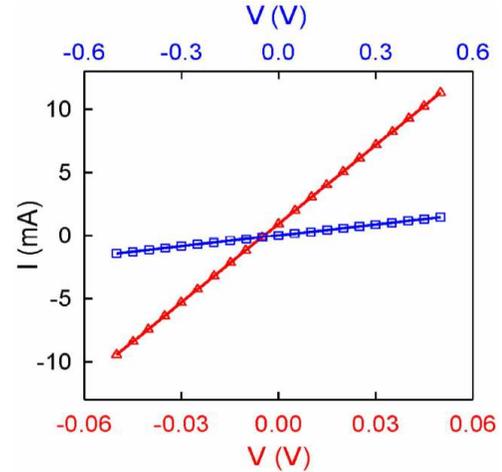} 
\caption{(Color online) Current voltage $ (I-V) $ characteristic curves for CS/PANi (red) and CS/PEO (blue) composites. The samples dimensions were similar. CS/PANi is more conducting than CS/PEO. Note that the voltage scale for CS/PANi is ten times smaller than for CS/PEO.
}
 \label{rateI}
\end{center}\end{figure} 

Figure 4(a) shows the temperature dependence of the conductivity of the CS/PANi and CS/PEO composites. The temperature dependence of pure PANi is also included for reference. Since PEO is an insulating polymer its contribution to the electrical conductivity is negligible and hence is not included. The temperature dependence of the conductivity of CS/PEO shows an initial increase as  temperature is lowered to 258 K (like a metal) below which it decreases (like a semiconductor). PANi and CS/PANi on the other hand show a monotonous decrease in the conductivity as temperature is lowered. PANi shows a steeper dependence on temperature compared to CS/PANi. By comparison, the temperature dependence of $\sigma $ for CS/PEO is weak. CS/PANi has the highest conductivity at room temperature and PANi taken alone the smallest one. In Figure 4(b) the conductivities are normalized to the values at 300 K to facilitate recognition of the different temperature dependences.  

\begin{figure}[t] 
\begin{center}
\includegraphics[width=11cm,height=7.8cm,angle=-90]{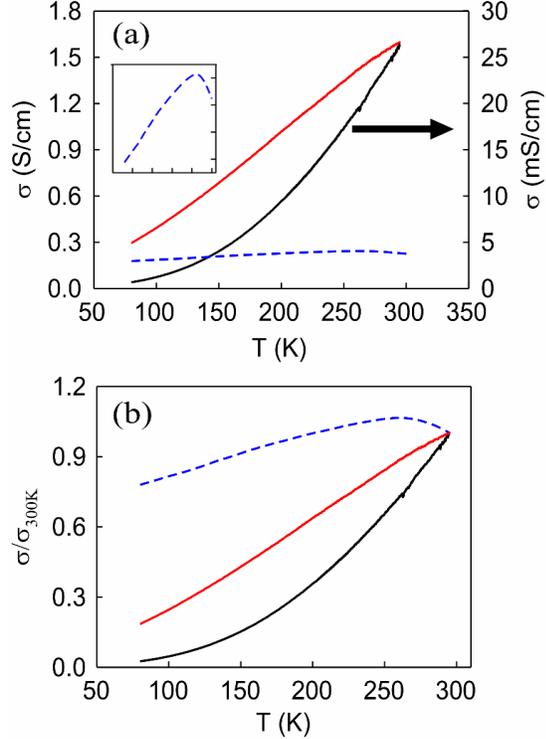} 
\caption{(a) Temperature dependence of the conductivity $(\sigma) $ of PANi (black – right scale), CS/PANi (red) and CS/PEO (blue). At room temperature, CS/PANi has the highest conductivity and PANi the lowest. A conductivity maximum was seen for CS/PEO at 258 K. PANi has a steeper temperature dependence compared to the composites. Inset: Temperature dependence of $ \sigma $  for CS/PEO seen in the main figure on an expanded scale. (b) Temperature dependence of the normalized conductivity $ (\sigma) $ of PANi (black – right scale), CS/PANi (red) and CS/PEO (blue). The conductivities are normalized to the values at 300K to facilitate recognition of the different temperature dependences.  
}
 \label{rateI}
\end{center}\end{figure} 

Figure 5(a) shows the natural logarithm of the conductivity plotted as a function of inverse temperature for all three samples, while Figure 5(b) shows the natural logarithm of the conductivity plotted as a function of $ T^{-1/4}. $ From these two figures, we see that neither thermal activation nor variable range hopping (VRH) models alone are controlling charge transport over the entire temperature range. We conclude that in each case, the temperature dependence of the conductivity is governed by the interplay of various charge transport mechanisms that include thermal activation, VRH and/or tunnelling, which are responsible for transport in different components of the considered composites (CS/PEO and CS/PANi).

\begin{figure}[t] 
\begin{center}
\includegraphics[width=11cm,height=8.3cm,angle=-90]{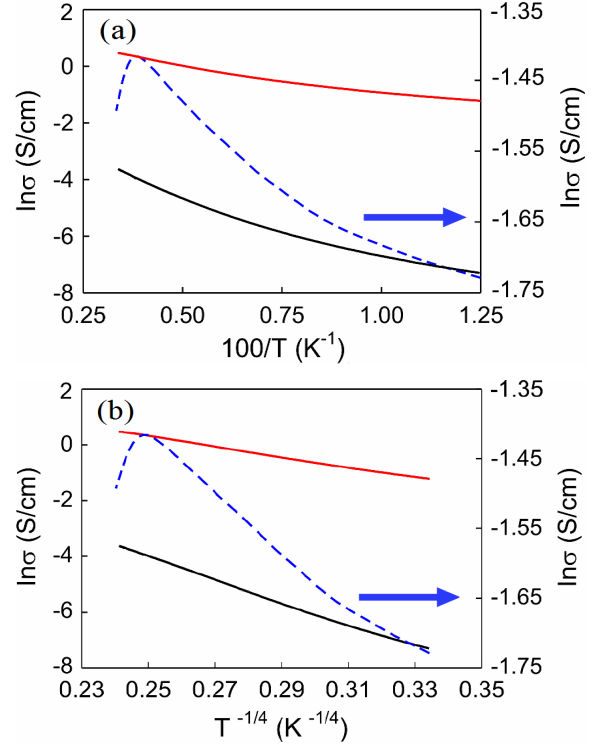} 
\caption{(a) Temperature dependence of the natural logarithm of the conductivity as a function of the inverse temperature. CS/PEO (blue -- right scale); CS/PANi (red); PANi (black).
(b)   Temperature dependence of the natural logarithm of the conductivity as a function of  $ T^{-1/4} $. CS/PEO (blue -- right scale); CS/PANi (red); PANi (black). 
}
 \label{rateI}
\end{center}\end{figure} 


{\bf 3b.} {\it Theoretical explanation:} In studies of electron transport through CS/PEO and CS/PANi composites, we assume that some carbon microspheres are in direct contact with each other. These CSs may be arranged in chains extended through the whole sample and maintaining pathways for traveling electrons as it occurs in self-assemblies of monodisperse starburst CSs \cite{9}. Electron transport along these pathways is mostly governed by the process of thermal activation typical for semiconducting materials. The corresponding contribution to the net conductivity is described by the commonly used Arrhenius relation:
\be
  \sigma_1 = \sigma_{10} \exp\left[-\frac{\Delta E}{k T} \right]           \label{1}
\ee
where $ \Delta E $   is the activation energy $ k $ is the Boltzmann’s constant, and the prefactor $\sigma_ {10} $  does not depend on temperature. Basing on the experimental data obtained for CS/PEO composites, it was found that $ \Delta E=2.3 meV $ \cite{9}.  A very close estimate $ (\Delta E=2.4 meV) $ was obtained for carbon fibers in polyethylene-carbon fiber composites and 
 for networked nano-graphites $ (\Delta E=4.0 meV) $ \cite{6,7}.  Here, we adopt the value  $ \Delta E=3.2 meV $ for the activation energy, which is the arithmetic mean between these two estimates. We stress that these estimations, solely concern CSs, and they are irrelevant for polymers included in CS/PEO and/or CS/PANi composites.

To compute the net conductivity of the  CS/PEO or CS/PANi sample, we mimic it by a couple of parallel channels. One of these channels represents all pathways through the chains of contacting carbon spheres. For this channel, electron transport is determined by thermal activation. Correspondingly, the conductance of this channel is proportional to the temperature dependent factor. We remark that due to the small value of $ \Delta  E $ this factor weakly depends on temperature especially at  $ T > 100 K. $ Therefore, the temperature dependence of the sample conductivity is mostly determined by the contribution from the second transport channel. This channel represents all remaining CSs which are separated by polymer filled gaps. 
  Electron transport through the gaps separating adjacent spheres may be controlled by two different transport mechanisms. These are the variable range hopping of electrons (VRH) between sites of polymeric chains and direct electron tunneling through the gaps. When VRH predominates, the resistivity is given by \cite{26}:  
\be
\rho = \rho_0 \exp  \bigg [\bigg( \frac{T_h}{T}\bigg)^{1/4} \bigg]  .  \label{2}
\ee
Here, $\rho_0 $  is independent on temperature, and the characteristic temperature $ T_h $ has the form:
\be
 T_h = \frac{16}{kN (E_F) l^3}  . \label{3}
\ee
In this expression, $ N(E_F) $ is the density of localized electron states for the  polymer and  $ l $ is the localization length for electrons which is proportional to the hopping distance $ R: $
\be
R = \frac{3}{8} l \bigg(\frac{T_h}{T} \bigg)^{1/4}  .  \label{4}
\ee
Usually, VRH transport mechanism predominates at  $T << T_h $ when the hopping distance significantly exceeds the localization length. For PANi \cite{27}, we estimate $ R $  and  $ l $ based on the data shown in the Fig. 6.  From the slope of the line representing PANi, we find the value of characteristic temperature $ T_h. $ Substituting this value into Eqs. (\ref{3}) and (\ref{4}) we obtain that $ l = 2.6  nm $ and $ R $ varies between $ 9.2 nm $ and  $13.1 nm $ as temperature decreases from 300 K to 80 K. Therefore, $ R > l $ over the whole range of temperatures used in the experiments.

So, one may expect VRH mechanism to significantly contribute to transport properties of CS/PANi composites. At the same time, PEO used in CS/PEO samples is an insulating polymer, and VRH mechanism does not noticeably affect transport characteristics of these samples. As for electron tunneling between the spheres, it occurs in both kinds of composites. However, we remark that in the case of conducting PANi filling space between CSs, electron tunneling occurs alongside VRH, whereas in the case of insulating PEO filling it solely governs electron transport between adjacent CSs \cite{9}.

The second transport channel may be regarded as a random resistor network where each resistor may be identified with a single sphere in series with a spacing filled with a polymer. The network conductivity/resistivity can be computed by the self-consistent effective medium theory \cite{28}. Within this approach one can replace all resistors in the network by the two resistors wired in parallel. In the first resistor, electron transport through PANi filled region is VRH controlled, and in the second one, electrons tunnel through this region. In the case of CS/PEO the whole network is reduced to a single resistor including a CS in series with a tunnel junction.

Figure 4(a) shows the temperature dependence of the conductivity $ (\sigma) $ for PANi, CS/PANi and CS/PEO.   Figure 4(b) shows the normalized composite conductivity as a function of temperature. The conductivity shown in Fig. 4(a) is much smaller than the composite conductivity, over the whole range of explored temperatures. Nevertheless, the PANi filled region contribution to the conductance may take on values of the same order or greater as the contribution from the CS. This happens because a typical spacing between the spheres $ (d \sim 10 nm - 100 nm) $ is much smaller than the mean CS diameter $  (D \sim 2 \mu m). $ The resistance is proportional to the sum $ \ds \frac{1}{\sigma_1} + \frac{\rho d}{D}$, where $ \sigma_1 $ and  $ \rho $ are given by Eqs. (\ref{1}) and (\ref{4}). At low temperatures, the second term in the sum strongly predominates. However, as the temperature rises, $ d $ rapidly ebbs and becomes comparable with 
$ D/\sigma_1. $ As a result, the corresponding contribution to the net conductivity 
\be
\sigma_2 = \frac{\alpha \sigma_1}{\ds \Big(1 + \frac{\rho\sigma_1 d}{D}\Big)}.   \label{5}
\ee
reveals a distinct temperature dependence although $ \sigma_1 $ levels off at higher temperatures.
In the Eq. (\ref{5}), the factor $\alpha $ describes the relative number of CS combined in the random network where electron transport through the gaps is governed by VRH with respect to the number of CS put in direct contact with each other. It is reasonable to assume that most CS are separated by gaps, so $\alpha $ takes on a value greater than 1.

Similarly, the contribution to the net conductivity from tunnel junctions may be presented
\be
\sigma_3 = \frac{\beta \sigma_1}{\ds \Big(1 + \frac{\sigma_1}{\sigma_t} \frac{d}{D}\Big)}.   \label{6}
\ee
Here, $ \sigma_t $ is the thermally induced tunnel conductivity of the form \cite{9}: 
\be
\sigma_t = \sigma_{30} \left(1 + \frac{T}{T_0} \right)^{-3/2} \exp\left[-\frac{T_1}{T + T_0}\right]    \label{7} 
\ee
where $\sigma_{30} $ is a temperature independent pre-factor, and characteristic temperatures  $T_1 $ and $ T_0 $ are given by relation \cite{28}:
\be
kT_1 = \frac{CV_0^2}{2} = \frac{\pi d \chi}{4} kT_0.   \label{8}
\ee

As follows from this relationship, values of $ T_1 $ and $ T_0 $ are determined by the height of the potential barrier separating adjacent CSs $ (V_0),$ by the capacitance  $C $ of a parallel plate capacitor representing the region where the tunneling between the spheres mostly occurs and by the tunneling constant $ \chi. $ Again, the factor $ \beta $ describes the relative number of tunnel junctions.
Summing up all contributions, one can present the net conductivity of the CS/PANi sample as:
\be
\sigma(T) = \sigma_1(T) + \sigma_2 (T) + \sigma_3 (T)  . \label{9}
\ee

\begin{figure}[t] 
\begin{center}
\includegraphics[width=7cm,height=5cm]{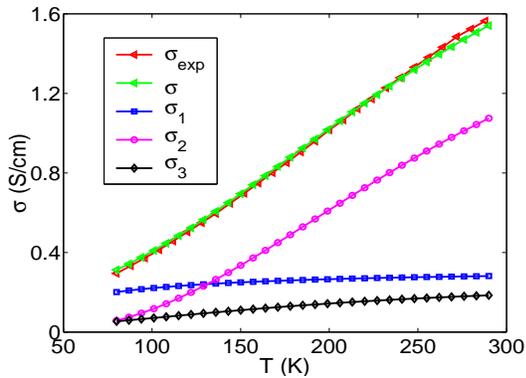} 
\caption{Theoretical results representing temperature dependent conductivity of CS/PANi composite. The curves are plotted using Eqs. (\ref{1}), (\ref{2}), (\ref{6}), (\ref{7}), (\ref{9}) assuming that $ \Delta E = 3.2 meV,\ \sigma_{10}  = 0.32 S/cm,\ \rho_0^{-1} = 380.8 S/cm,\ \sigma_{30} = 18 S/cm,\ T_h = 2.5 \times 10^6 K,\ T_0  = 200 K,\ T_1 = 1850 K,\ d/D = 0.075.$ The parameters $ \alpha $  and $ \beta $ are found to be $ \alpha = 7.1 $ and $ \beta = 1 $ by employing the least mean squares method to provide the best fit between theory and experimental data.
}
 \label{rateI}
\end{center}\end{figure}

The temperature dependences of the net conductivity $\sigma (T) $ and of  $ \sigma_1(T), $ of $ \sigma_2(T), $ and $\sigma_3(T) $ are displayed in Figure 6. The fit with the experimental data shows that our model for charge transport in CS/PANi is valid over the whole temperature range under study. One observes that at low temperatures $ (T < 120 K) $ the conductivity is mostly governed by thermally activated transport through carbon spheres. At higher temperatures  $(120 K < T < 300 K) $ the VRH transport mechanism predominates. As for thermally induced tunneling, its contribution remains smaller than the other two but it cannot be neglected, especially for higher temperatures. Therefore, all three transport mechanisms combined together determine the quasilinear temperature dependence of CS/PANi sample observed in experiment. In CS/PEO composites VRH transport is negligible, and the temperature dependence of the sample conductivity is controlled by the interplay between thermally activated transport through chains of contacting CS and thermally induced tunneling through PEO filled gaps. 

 Ideally, it is preferable to have a composite with high conductivity like CS/PANi and weak temperature dependence like CS/PEO. Polyaniline doped with camphor sulfonic acid (CSA) and dissolved in m-cresol was shown to have high conductivity and a weak temperature dependence compared to that doped with HCl [26]. Further work to prepare CSs composites with CSA doped PANi should increase the conductivity even more and yield a weaker temperature dependence. This will have the advantage of incorporating the desirable characteristics of both types of composites studied above. Experiments to verify this result are currently being planned.
 
\section{4. Conclusions}

Carbon spheres  with diameters in the range $ 2 \mu m-10 \mu m $ were prepared via hydrolysis of a sucrose solution at $ 200^o C $ and later annealed in $ N_2 $ at $ 800^o C $. Composites of these spheres with PEO and with PANi were prepared and the charge transport studied in the temperature range of   $ 80 K < T < 300 K. $  The CS/PANi composites exhibited higher conductivity compared to CS/PEO in the entire temperature range. The conductivity in CS/PEO showed metallic-like behavior down to 258 K, below which the conductivity decreased, while the conductivity in CS/PANi decreased monotonically when the temperature decreased over the entire temperature range. We have analyzed possible transport mechanisms in CS/PEO and CS/PANi composites assuming that some part of carbon spheres are glued together forming chains which may be extended through the whole sample. Another (and, probably, greater) part of CSs are separated from each other by polymer filled gaps. The chains of CSs are organized as a network of channels for thermally-activated electron transport (typical for intrinsic semiconductors) with low activation energy found from experiments on polyethylene-carbon fiber, CS/PEO composites and networked nano-graphites \cite{6,7,9}.  The charge transport through CS separated by gaps is controlled by other transport mechanisms. In the case of CS/PEO, the polymer filling these gaps is nearly insulating, so electrons can travel from one sphere to another one solely by means of direct thermally induced tunnelling. In CS/PANi samples the gaps are filled with a conducting polymer where VRH mechanism governs the charge transport. So, in CS/PANi composites electrons may either tunnel between adjacent CSs or travel through the PANi filled gaps by means of VRH. It follows from our experiments that the latter mechanism predominates in charge transport through PANi filled gaps. Finally, we conclude that charge transport in CS/PANi occurs due to the combined effects of 3-D variable range electron hopping, and tunnelling between CSs combined with thermally activated transport through the chains made from contacting CSs,  while that in CS/PEO transport, was due to tunnelling and activation. The high conductivity in both composites make these materials attractive in the fabrication of devices and sensors.
\vspace{2mm}

{\bf Acknowledgements:} This work was supported in part by NSF under grants: DMR-PREM 1523463 and DMR-RUI 1360772.  The authors are grateful to G. M. Zimbovsky for help with the manuscript.

\end{document}